# AutoMoDe –Model-Based Development of Automotive Software

Dirk Ziegenbein
Robert Bosch GmbH

Ulrich Freund
ETAS Engineering Tools GmbH

Peter Braun
Validas AG

Andreas Bauer, Jan Romberg, Bernhard Schätz
Institut für Informatik, TU München

Contact: dirk.ziegenbein@de.bosch.com

## Abstract

*This paper describes first results from the AutoMoDe (Automotive Model-Based Development) project. The overall goal of the project is to develop an integrated methodology for model-based development of automotive control software, based on problem-specific design notations with an explicit formal foundation. Based on the existing AutoFOCUS framework [1], a tool prototype is being developed in order to illustrate and validate the key elements of our approach.*

## 1 Introduction

Traditionally, the focus of automotive software engineering is on the later and more detailed abstraction levels, which deal strongly with implementation-related issues. For lack of suitable notations, methodologies, and integration between abstraction levels, more abstract system descriptions typically take a back seat in the design process. However, working at higher levels of abstraction will be a key factor in tackling the prevalent complexity issues in automotive software engineering:

- increasing functional complexity stemming from software being the implementation technology of choice for new innovative functionality as well as for functionality being traditionally implemented in mechanics etc.
- complex relations between design artifacts, such as large number of variants in product families
- design chains spanning several technical disciplines and organizations/companies

The impact of this complexity has led to the start of several projects trying to define methods and tools to raise the development of embedded automotive control software to higher abstraction levels. For example, a decade ago DaimlerChrysler started to tailor ROOM-inspired concepts for body electronic systems in the TITUS project [2]. In 1998, the French automotive industry started the AEE project [3]. Both projects resulted in a common European effort called EAST/EEA [4] providing as one of the results an automotive architecture description language (ADL) [5]. Currently, the AUTOSAR development partnership [6] is up to ease the development of automotive software mainly through providing a standardized infrastructure and methodology.

As a result of these efforts, modeling means providing a lot of feasible abstraction mechanisms are available. Unfortunately, these modeling means cover only some aspects of embedded automotive software design like networks, control-algorithms, or software architecture, while an accepted mature modeling framework is still missing.

A major requirement for such a modeling framework is the provision of several system abstractions tailored for different stakeholders and different phases in the design process. A methodology should provide support for easy transitions between these abstraction levels. Such transitions should be preferably tool-supported. Notations and underlying models have to be well-integrated to ensure consistency between different abstractions which is crucial for a design process typically spanning several companies.

In order to provide a modeling framework with tailored abstraction levels, a well-defined operational model, and formalized transformation steps, we propose the model-based approach AutoMoDe - Automotive Model-based Development.

The paper is organized as follows. Sec. 2 introduces the AutoMoDe operational model with explicit data-flow and discrete-time semantics. The different abstraction levels of AutoMoDe and its graphical notations are presented in Sec. 3. Sec. 4 introduces a classification of transformation steps, intended to ease optimizing models and bridging between system abstractions. Sec. 5 discusses experiences made during a reengineering case study using the AutoMoDe approach. Finally, Sec. 6 concludes the paper and gives an outlook on future work.

## 2 Operational Model

The operational model of AutoMoDe is an extended version of the established model of the AutoFOCUS framework [1], which features a message-based, discrete-time communication scheme as its core semantic model. Thus, each AutoMoDe model element can be understood as a component or block exchanging messages with its environment via logical channels with respect to a global, discrete time-base. This computational model supports a high degree of modularity (by making component interfaces complete and explicit) as well as a reduced degree of complexity (by abstracting from implementation details such as detailed timing or communication mechanisms):

- The message-based communication with explicit data-flow enforces complete specification of a component's interface, and prohibits implicit exchange of information, such as undocumented access of global variables.
- The discrete-time communication avoids the use of timing assumptions below the chosen granularity of observable discrete (abstract) clock ticks. Real-time intervals of the implementation are therefore abstracted by logical time intervals. Within these logical time intervals, no assumptions on the ordering of message arrivals or on duration and delays of message transfers are made.

Note that this communication model does allow for both periodic and sporadic communication as required for a mixed modeling of time-triggered and event-triggered behavior. As shown in Figure 1, each channel in the abstract model either holds an explicit value or the "-" ("tick") value indicating the absence of a message. Thus, by reacting explicitly depending on the presence (or absence) of a message, modeling of event-triggered behavior is naturally covered by the AutoMoDe description techniques.

One enhancement of AutoMoDe in comparison to the AutoFOCUS operational model is the explicit support of multi-rate systems, i.e. systems featuring signals with different frequencies. Each message flow in AutoMoDe is associated with an abstract *clock*. For a given flow, the flow's clock indicates either the frequency of message exchange (periodic case), or a condition describing the event pattern (aperiodic case). Syntactically, a clock is simply a Boolean expression evaluating to logical "true" whenever a message is present on the clock's flow.

Specific operators such as "delay" or "when" allow for well-defined sampling within a model of different abstract clocks, i.e. signal frequencies. This concept of sampled clocks originates from the field of synchronous languages [7].

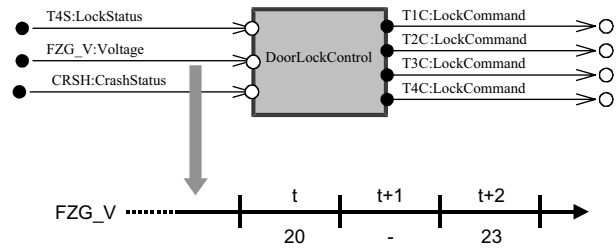

Figure 1. Message-Based, Time-Synchronous Communication

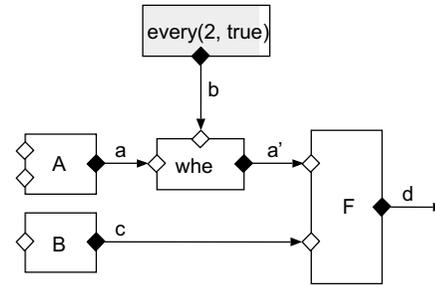

Figure 2. Explicit signal sampling using a "when"-operator.

In the example depicted in Fig. 2, a "when" operator is used to sample the stream of signals *a* down by a factor of two. This factor is denoted by the Boolean clock expression "every(2, true)". "every(n, true)" is a macro operator which yields a "true" value each $n^{th}$ tick of the base clock (always "true"), and a "false" value otherwise. Consequently, *a'* is updated every second tick of the base clock.

Obviously, the combination of a globally clocked operational model with distributed automotive E/E architectures featuring event-triggered, not tightly synchronized communication media such as the CAN bus poses some research questions. In [8], a proposal is presented on how to use event-triggered media for firm real-time deployment of globally clocked models with comparatively small implementation overhead. However, this topic will be subject of further investigation.

## 3 Abstraction Levels and Notations

Of central importance for the model-based approach of AutoMoDe are the different system abstractions visible to the designers and their supported views on the system (see Fig. 3) as they determine the usability of the approach. While the system abstractions are specifically targeted to certain tasks and stakeholders in the design process, the information offered in these views are abstracted from the coherent AutoMoDe meta-model of the system. Thus, consistency between abstraction levels is guaranteed.

The chosen system abstraction levels are similar to those defined in [5], but are adapted to match the model-based AutoMoDe development process. In the following, these system abstraction levels and their respective AutoMoDe notations are introduced.

### 3.1 Functional Analysis Architecture

The Functional Analysis Architecture (FAA) is the most abstract layer considered in AutoMoDe. The FAA provides a system-level abstraction representing the vehicle functionalities to be implemented in either hardware or software. The FAA addresses the integration of separately developed vehicle functions on the functional level, i.e. implementation details and qualitative requirements are not considered. This allows for a system representation relatively easy to understand and targeted at function developers and customers.

An FAA-level description is typically complete with respect to the considered functionalities, and the functional dependencies between them. It is then possible to identify functional dependencies and potential conflicts between vehicle functions, and the validation of functional concepts based on prototypical behavioral descriptions. Means to achieve these goals include rules as well as model simulation. Based on the functional structure and dependencies, rules identify possible conflicts (e.g. two vehicle functions access the same actuator) and suggest suitable countermeasures to resolve them (e.g. introduce a coordinating functionality). The simulation additionally considers the prototypical behavioral descriptions. These descriptions are not optimized for efficient implementation and abstract from details such as concrete data types.

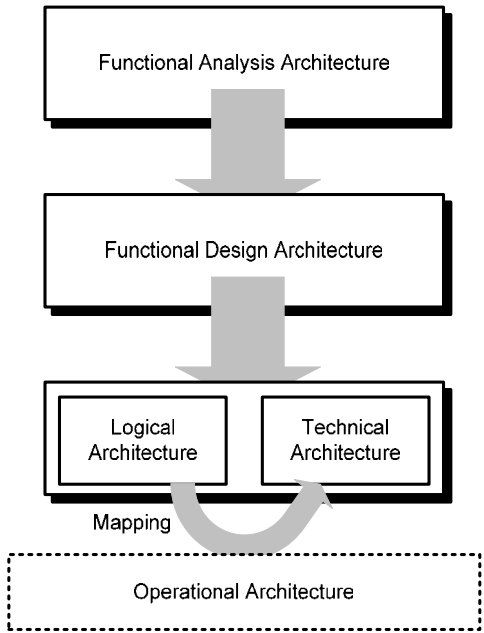

Figure 3. AutoMoDe Abstraction Levels.

**System Structure Diagrams**

The dominating notation used on the FAA level is called System Structure Diagram (SSD). SSDs are used for describing high-level architectural decomposition of a system, similar to UML 2.0 component diagrams. SSDs consist of a network of typed *components* with statically typed message-passing interfaces (*ports*). Explicit connectors (*channels*) connect ports and indicate the direction of message flow between components. Components can be either recursively defined by other SSDs, or by different notations for behavioral description (Sec. 3.2). On the FAA level, it may be perfectly adequate to leave the detailed behavior unspecified. For an example SSD, see Fig. 4.

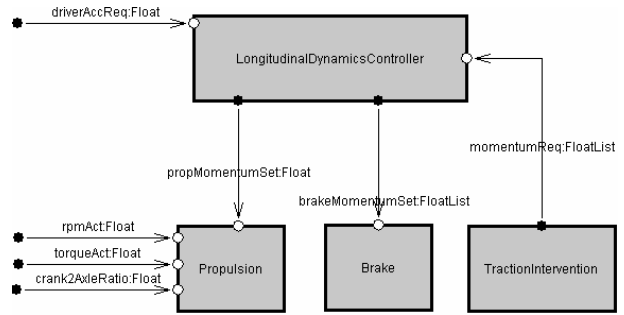

Figure 4. System Structure Diagram (SSD) on the FAA level.

The component boundaries introduced by SSDs have semantic implications as well – each SSD-level channel introduces a message delay. Because of AutoMoDe's global discrete-time semantics, such implicit introduction of delays is done to facilitate later design transformations such as deployment.

Note that SSDs are not unique to the FAA, but will be used on other abstract system levels as well (see following sections).

### 3.2 Functional Design Architecture

The AutoMoDe system abstraction Functional Design Architecture (FDA) is a structurally as well as behaviorally *complete* description of the software part in terms of actual software components that can be instantiated in later phases of the development process. For coarse-grained decomposition of the design, again SSDs are used.

In contrast to FAA-level functionalities, the FDA-level software components are formed in order to satisfy qualitative requirements such as portability, performance, maintainability, reuse, etc. Thus, the FDA is targeted to software architects as well as for individual components to software developers.

On FDA level, the atomic SSD are required to have a well-defined behavior which can be specified using the following three AutoMoDe notations.

**Data Flow Diagrams**

Data Flow Diagrams (DFD) define an algorithmic computation of a component. Graphically, DFDs are similar to SSDs (see Fig. 5): DFDs are built from individual *blocks* with dynamically typed ports connected by channels. Blocks may be recursively defined by another DFD; the behavior of atomic DFD blocks is then given either through a Mode Transition Diagram (MTD), a State Transition Diagram (STD), or directly through an expression (*function*) in AutoMoDe's base language. For example, block ADD in Fig. X is defined by the function ch1+ch2+ch3. With this mechanism, it is possible to define adequate block libraries for discrete-time computations.

The default semantics of DFD communication is "instantaneous" in the sense of synchronous languages [7]. In the AutoMoDe tool prototype, instantaneous communication primitives are accompanied by a causality check for detecting instantaneous loops. Note that computations "happening at the same time" in the FAA., FDA- or LA-level models are perfectly valid abstractions of sequential, time-consuming computations on the level of the Operational Architecture (OA) if the abstract model's computations are observed with a delay, such as the delays introduced by SSD composition. The duration of the delay then defines the deadline for the sequential computation in the OA.

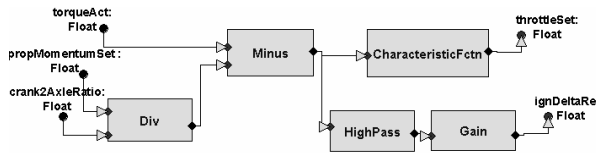

Figure 5. An AutoMoDe DFD representing a longitudinal momentum controller.

**Mode Transition Diagrams**

In order to represent explicit system modes and alternate behaviors w.r.t. modes, Mode Transition Diagrams (MTDs) are used. MTDs consist of *modes*, and *transitions* between modes (see Fig.6). Transitions are triggered by certain combinations of messages arriving at the MTD's component. The behavior of the component within a mode is then defined by a subordinate DFD or SSD associated with the mode (comparable to the composition of FSMs and concurrency models in *charts [9]). As illustrated by the example in Sec. 5, MTDs provide a valuable means of architectural decomposition specifically suited for embedded control systems.

The usage of explicit operational modes for architecture-level decomposition has also been brought forward by other authors [e.g. 10]. In addition to the basic idea of using explicit notations for operational modes, our approach focuses on the use of mode representations spanning several abstraction levels and on transformations

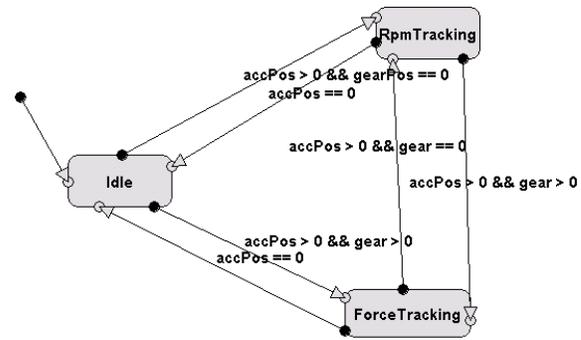

Figure 6. An AutoMoDe MTD specifying engine operation modes.

between different mode representations suited for different abstraction levels.

**State Transition Diagrams**

State Transition Diagrams (STDs) are extended finite state machines similar to the popular Statecharts notation, but with some syntactic restrictions for excluding certain semantic ambiguities allowed by some standard Statecharts dialects [11].

### 3.3 Logical and Technical Architecture

The Logical and Technical Architecture (LA, TA) is the most implementation-oriented abstraction level supported by the AutoMoDe approach. The LA mainly groups and instantiates FDA-level components to *clusters*. The TA represents target platform components (ECUs, tasks, buses, message frames) used to implement the system.

The LA/TA abstraction level is targeted to system architects and provides all means necessary to defining the deployment of SW components to the target platform.

A cluster can be thought of as a "smallest deployable unit". Consequently, several clusters may be mapped to a given operating system task, but a given cluster will not be split across several tasks.

**Cluster Communication Diagrams**

The notation used for top-level definition of the LA structure is called Cluster Communication Diagrams (CCD). An example is depicted in Fig. 7. Like SSD components, clusters have statically typed interfaces – moreover, signal frequencies are made explicit on the LA level. In contrast to SSDs and DFDs, Clusters may not be defined recursively by other CCDs (but hierarchical DFD descriptions are perfectly adequate for defining the internal behavior of clusters). The graphical representation of CCDs is similar to DFDs.

When transitioning from an SSD representation on the FDA level to a LA-level CCD, some of the topmost SSD hierarchies may be dissolved in favor of a flat CCD representation. In order to represent high-level MTDs as a network of clusters on the LA level, the AutoMoDe tool prototype features an algorithm to transform an MTD into a semantically equivalent, partitionable data-flow model.

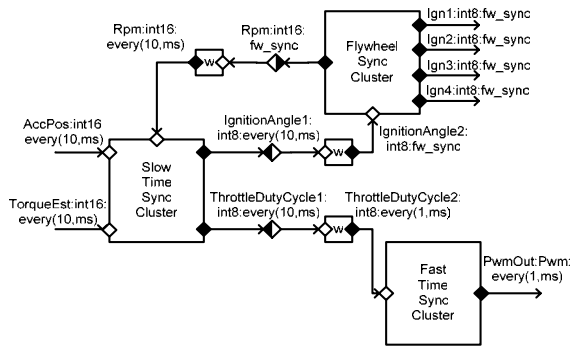

Figure 7. An AutoMoDe CCD representing a simplified engine controller.

The type system at the LA level is extended by *implementation types* which capture the platform-related constraints associated with implementation. That means, abstract data types such as `int` are typically mapped to implementation, e.g. `int16` or `int32`. Similarly, a floating-point message on the FDA level may be mapped to a fixed-point or integer message on the LA level.

For CCDs, well-definedness conditions can be specified that may depend on the characteristics of a given Technical Architecture. As an example, consider an OSEK-conformant operating system as a target platform, with inter-task communication between tasks using data integrity mechanisms [12] and fixed-priority, preemptive scheduling. In this framework, communication from "slower-rate" clusters to a "faster-rate" cluster necessitates the introduction of at least one delay operator in the direction of data flow. On the other hand, communication in the opposite direction ("fast-rate" to "slow-rate" cluster communication) does not require introduction of delays in the CCD. Consequently, CCD well-definedness conditions may be adapted to the specific target architecture considered for implementation.

### 3.4 Operational Architecture

The result of the deployment of SW clusters to the target architecture is the starting point of the Operational Architecture (OA). However, this abstraction level is not part of the AutoMoDe tool prototype as there is already commercial tool support for this level of abstraction, e.g. ASCET-SD [13]. Thus, based on the deployment decisions, the AutoMoDe tool prototype will generate ASCET-SD projects for each ECU of the target architecture.

All signals between clusters deployed to different ECUs will be mapped to a communication network, e.g. CAN, possibly considering an existing communication matrix. In all generated ASCET-SD projects, additional communication components have to be added which can be configured according to the generated or supplemented communication matrix.

### 4 Transformations

Besides adequate modeling means, the core of the AutoMoDe approach is the investigation of and tool support for model transformations. Three different types of transformation steps are considered.

As stated above, automotive control software is rarely developed from scratch. *Reengineering* is seen as the step to extract the relevant information from a system description on the implementation level in order to describe the system on a more abstract level (FAA or FDA). Two classes of reengineering steps are considered. While "white-box" reengineering considers complete software implementations (e.g. ASCET-SD models) of functions (see case study in Sec. 5), "black-box" reengineering transforms E/E architecture representations like communication-matrices, which capture dependencies between functions, to partial FAA level representations. This "black-box" reengineering approach is currently being validated with a body-electronics case study.

*Refactoring* is mainly seen as a structural transformation on the same abstraction level. An example is the integration of an independently designed control algorithm into an FAA-level functional network. The algorithm has to be restructured considerably because e.g. other functions access the same actuator such that the structural hierarchy of the control algorithm has to be adapted. Other refactoring steps will replace an MTD by several DFDs having explicit mode-ports, or change the structural hierarchy in order to facilitate more efficient implementation.

*Refinement* is the transformation from higher to lower abstraction levels. Examples for refinement transformations include the transformation of physical signals to implementation signals (i.e. the choice of encoding and data type), clustering of DFDs according to their clocks neglecting their functional coherency and last but not least the mapping of CCDs to ECUs and tasks.

### 5 Case Study

The above concepts (see Sec. 2 – 4) have been applied to an extensive automotive case study of a four-stroke gasoline engine control algorithm. This case study was provided in terms of a detailed ASCET-SD model and has been reengineered in significant parts using a first AutoMoDe tool prototype along with the according notations and underlying semantics.

Compared to ASCET, AutoMoDe provides a richer set of control flow primitives. As it turns out, the AutoMoDe notion of modes and MTDs is able to capture and encapsulate implicit operation modes of the original ASCET model. More so, implicit modes of ASCET processes can be made explicit to the developer by using MTDs, rather than control flow operators such as If-Then-Else (see Fig. 8).

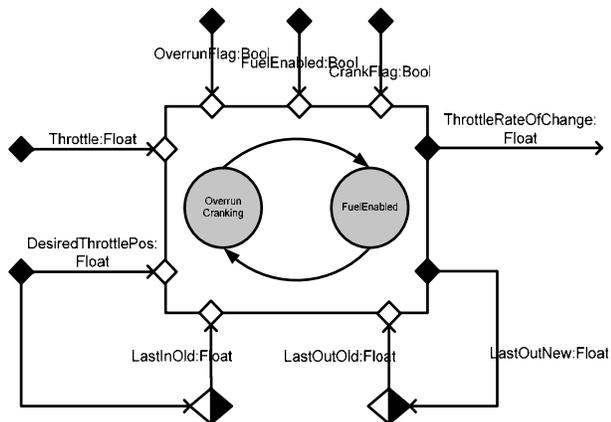

Figure 8: AutoMoDe component with an embedded MTD which consists of two states: "FuelEnabled" and "CrankingOverrun".

For example, a component "ThrottleRateOfChange" determines the change rate of the throttle valve position not only depending on its current and the desired position, but also depending on very specific states of the entire engine. More traditional approaches would suggest to use conditional operators such as If-Then-Else to either respond with a constant factor or to trigger a more complex algorithmic computation for a more detailed determination of the change rate. Modeling "ThrottleRateOfChange" with modes, however, divides the component in two states which are being modeled and viewed separately, depending on the respective engine state (see Fig. 8).

The introduction and use of modes in an AutoMoDe models increases the consistency of the model by making orthogonal modes explicit. This became strikingly apparent in the case study where in the original model a centralized software component emits a large number of flags which altogether represent the global state of the engine. Due to the high complexity of this central component, it is unclear which disjunctive states or modes exist at all, let alone isolate the model parts which are active in a certain mode.

In contrast, MTDs offer a conceptually clear way to represent state explicitly, rather than relying on implicit control-flow dependencies. Moreover, the different modes in MTDs can be used in order to determine a global mode transition system which is then correct by construction.

The reengineered model being constructed in this case study will be used to evaluate refactoring and refinement steps in future.

## 6    Conclusion

The AutoMoDe approach combines the advantages of having a thorough and consistent operational model with the existence of well-defined system abstractions specifically targeted to typical tasks in automotive software development.

In order to support the typically costly and error-prone tasks of bridging between abstraction levels or optimizing system models, identified transformation steps will be formalized and supported by the AutoMoDe tool prototype.